\documentclass[pra,twocolumn,showpacs,floatfix,superscriptaddress,aps]{revtex4}
\usepackage{amsmath}
\usepackage{amssymb}
\usepackage{graphicx}
\usepackage[usenames]{color}
\usepackage{dcolumn}
\begin{document}

\title{Ultracold collisions between two light indistinguishable diatomic molecules: 
elastic and rotational energy transfer in HD+HD}

\author{Renat A. Sultanov\footnote{rasultanov@stcloudstate.edu
 (r.sultanov2@yahoo.com)}}
\affiliation{Instituto de F\'isica Te\'orica, UNESP $-$ Universidade Estadual Paulista, 
01140 S\~ao Paulo,
SP, Brazil}
\affiliation{Department of Information Systems and BCRL,
St. Cloud State University, St. Cloud, MN, USA}

 \author{Dennis Guster\footnote{dcguster@stcloudstate.edu}}
\affiliation{Department of Information Systems and BCRL,
St. Cloud State University, St. Cloud, MN, USA}

\author{S. K. Adhikari\footnote{adhikari@ift.unesp.br; 
http://www.ift.unesp.br/users/adhikari}}
\affiliation{Instituto de F\'isica Te\'orica, UNESP $-$ Universidade Estadual Paulista, 
01140 S\~ao Paulo,
SP, Brazil}

\date{\today}

\begin{abstract}
A close coupling quantum-mechanical calculation is performed for
rotational energy transfer in a HD+HD collision at very low energy, down to
the ultracold temperatures: $T \sim 10^{-8}$ K.
A global six-dimensional H$_2$-H$_2$ potential energy surface  
is adopted from a previous work [Boothroyd {\it et al.}, J. Chem. Phys., {\bf 116},  666 (2002).]
State-resolved integral cross sections $\sigma_{ij\rightarrow i'j'}(\varepsilon_{kin})$
of different quantum-mechanical rotational transitions $ij\rightarrow i'j'$
in the HD molecules and corresponding state-resolved thermal rate coefficients
$k_{ij\rightarrow i'j'}(T)$ have been computed. Additionally, for comparison,
H$_2$+H$_2$ calculations for a few selected rotational transitions have also been performed.
The hydrogen and deuterated hydrogen molecules are treated as rigid rotors in this work.
A pronounced isotope effect is identified in the cross sections of these collisions at low
 and ultracold temperatures.
\end{abstract}


\maketitle

\section{Introduction}
The recent  creation and investigation of a quantum gas of ultracold diatomic  molecules
\cite{ni08} is of great interest in many areas of atomic, molecular, optical, and
chemical physics \cite{demille02,rabl06,hudson08,hershbach09,bell09}.
Research in these fields
may have important future applications, for example, in quantum information processing 
\cite{zoller05,polzik10,mishima11,gorshok11}. From a scientific point of view the creation of the molecular quantum
gas opens new doors, for instance, in the experimental and theoretical investigation of the cold and
ultracold molecular scattering and
chemical reactions \cite{hutson07,lee2006,krems08,sawyer08,roudnev09p,renat2011a}. It allows researchers to probe
the interaction and collisional properties of different light and heavy  molecules in the cold
and ultracold regime:
$T \sim 10^{-4} - 10^{-8}$ K
\cite{ni10,ni09,hutson07,hershbach09,bell09}.
In this regime, one can expect many shape resonances in the cross
sections arising from the van der Waals force \cite{hershbach09, bell09}. 
For example, a resonance with a weakly
bound level near zero collision energy can significantly enhance the tunneling effect through a
reaction barrier. By aligning and orienting the colliding molecules, the anisotropy of the van der Waals
forces enables substantial tuning of the molecular levels to create such resonances \cite{hershbach09}.

In this work, the ultracold collision between two deuterated hydrogen molecules, i.e. rotational
energy transfer in HD+HD, is mainly considered. For comparison similar collision 
in  H$_2$+H$_2$ is also considered.
From a theoretical point of view the HD+HD system is interesting because
its PES can be derived from the much-studied H$_2$+H$_2$
system by adjusting the coordinate of the HD-molecule center of mass.
Once the symmetry is broken in H$_2$-H$_2$ by replacing the H with a D atom
in each H$_2$ we have the precise HD-HD PES.  
The HD and H$_2$ molecules are treated as rigid monomer rotors in this work, so  we ignore 
the vibrational degrees of freedom of these molecules.

Because of the small reduced mass and large rotational-energy spacing in the HD-HD system,
the number of states required in the basis set for an accurate quantum-mechanical
calculation should be relatively small. The HD+HD system has 
widely spaced rotational-energy levels and, because of the strong anisotropy of the intermolecular potential, it
has relatively large rotational-energy transition probabilities.
Since HD is a light molecule it can be  manipulated easily by an external electrical field and
also, the laser cooling  of this diatomic molecule seems possible making  this system is of current experimental interest. 

Surprisingly, such a fundamental and attractive quantum four-atomic system has not
received substantial attention in previous experimental and theoretical investigations.
Several molecular-beam studies of HD+HD involve the
measurement of a few rotational probabilities \cite{gentry77},
integral cross sections for unresolved internal \cite{johnson79},
and rotational energy transfer rates
\cite{chandler86,chandler88}. Nevertheless, there are  only a few
calculations dealing with the rotational excitation in the HD+HD collision, e. g., an early
modified-wave-number calculation by Takayanagi \cite{taka59}, semiclassical calculations
by  Gelb and Alper \cite{gelb79}, and Cacciatore and Billing \cite{billing92}.

Hydrogen isotope effects have
often attracted considerable attention \cite{kusakabe04}. In this work we carry out such
consideration within the HD+HD and H$_2$+H$_2$ systems at high, low and ultracold temperatures.
In the next section we  briefly present the quantum-mechanical approach, that is used in this
work and the PES. In Sec. III we present  numerical results for both HD+HD and H$_2$+H$_2$
collisions. Additionally, we present a brief discussion of the numerical convergence
of the results. Finally, in Sec. IV we present a summary and  conclusion.

\section{Method: quantum dynamics}

\begin{figure}
\begin{center}
\includegraphics[width=.92\linewidth,clip]{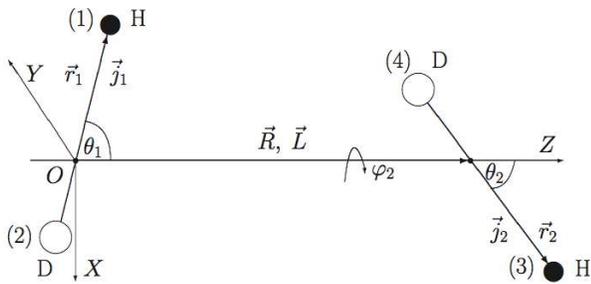}    
\caption{Four-atomic system $(12)+(34)$ or HD+HD, where H is a hydrogen atom
and D is deuterium, represented by few-body Jacobi coordinates: $\vec r_1$, $\vec r_2$, and $\vec R$.
The vector $\vec R$ connects the center of masses of the two HD molecules and
is directed over the axis $OZ$, $\theta_1$ is the angle between $\vec r_1$ and $\vec R$,
$\theta_2$ is the angle between $\vec r_2$ and $\vec R$, $\varphi_2$ is
the torsional angle, $\vec j_1, \vec j_2$, and $\vec L $ are quantum angular momenta over the corresponding
Jacobi coordinates $\vec r_1$, $\vec r_2$, and $\vec R$.}
\label{fig1} \end{center} \end{figure}

Here we  briefly  present the close-coupling quantum-mechanical approach used in this study 
to calculate the cross sections and thermal rate coefficients of molecular 
hydrogen-hydrogen collision.
The Schr\"odinger equation for the  $(12)+(34)$ collision in the center of a mass frame,
where $(12)$ and $(34)$ are linear rigid rotors is \cite{green75,green78}:
\begin{eqnarray}
\left(\frac{P^2_{\vec R}}{2M_{12}}+\frac{L^2_{\hat r_{1}}}{2\mu_1r_1^2}+\frac{L^2_{\hat r_{2}}}{2\mu_2r_2^2}+
V(\vec r_1,\vec r_2,\vec R) - E \right)\times \nonumber \\
\Psi(\hat r_{1},\hat r_{2},\vec R)=0,
\label{eq:schred}
\end{eqnarray}
where $P_{\vec R}$ is the relative momentum operator, 
$\vec R$ is the relative position vector,
$M_{12}$ is the reduced mass of the pair  
%
$M_{12} = (m_1+m_2)(m_3+m_4)/(m_1+m_2+m_3+m_4)$,
%
$\mu_{1(2)}$ are reduced masses of the targets:
%
$\mu_{1(2)}=m_{1(3)}m_{2(4)}/(m_{1(3)}+m_{2(4)})$,
%
$\hat r_{1(2)}$ are the angles of orientation of rotors $(12)$ and $(34)$, respectively,
%
%
$E$ is the total center-of-mass energy and
$V(\vec r_1,\vec r_2,\vec R)$ is the potential energy surface for the four atomic system $(12)+(34)$.
The system is shown in Fig. \ref{fig1}.
Basically, the PESs of the H$_2$-H$_2$ and the HD-HD systems 
are the same. However, there is a small but important difference. 
To obtain the HD-HD PES from the existing H$_2$-H$_2$ surface \cite{booth02} one needs to 
appropriately shift the center of mass in the hydrogen molecules (H$_2$).
The usual rigid rotor model \cite{green75,schaefer90,flower98,renat06,renat10}
has also been applied in astrophysical calculations
of different atom and diatomic-molecule collisions
or two diatomic-molecule collisions at low temperatures: $T < 2000$ K.

The eigenfunctions of the operators $L_{\hat r_{1(2)}}$ 
in Eq. (\ref{eq:schred}) are simple spherical harmonics $Y_{j_im_i}(\hat r)$. To solve
Eq. (\ref{eq:schred}) the following expansion is used \cite{green75}:
\begin{eqnarray}
\Psi(\hat r_{1},\hat r_{2},\vec R)=\sum_{JMj_1j_2j_{12}L}\frac{U^{JM}_{j_1j_2j_{12}L}(R)}{R}\nonumber \\
\times \phi^{JM}_{j_1j_2j_{12}L}(\hat r_1,\hat r_2,\vec R),
\label{eq:expn}
\end{eqnarray}
where $J$ is the total angular momentum quantum number, $M$ is its projection onto the space fixed $z$ axis
and the channel expansion functions are
\begin{eqnarray}
\phi^{JM}_{j_1j_2j_{12}L}(\hat r_1,\hat r_2,\vec R) = \sum_{m_1m_2m_{12}m}C_{j_1m_1j_2m_2}^{j_{12}m_{12}}
C_{j_{12}m_{12}lm}^{JM}\nonumber \\
\times Y_{j_1m_1}(\hat r_1)Y_{j_2m_2}(\hat r_2)Y_{Lm}(\hat R),
\end{eqnarray}
with $j_1+j_2=j_{12}$, $j_{12}+L=J$, $m_1$, $m_2$, $m_{12}$ and $m$ the projections of $j_1$, $j_2$, $j_{12}$ 
and $L$ respectively.

Substitution of Eq. (\ref{eq:expn}) into (\ref{eq:schred}) provides a set of coupled second order differential equations
for the unknown radial functions $U^{JM}_{\alpha}(R)$
\begin{eqnarray}
\left(\frac{d^2}{dR^2}-\frac{L(L+1)}{R^2}+k_{\alpha}^2\right)U_{\alpha}^{JM}(R) = \nonumber \\
2M_{12}\sum_{\alpha'}\int <\phi^{JM}_{\alpha}(\hat r_1,\hat r_2,\vec R)
|V(\vec r_1,\vec r_2,\vec R)|\nonumber \\
\phi^{JM}_{\alpha'}(\hat r_1,\hat r_2,\vec R)>U_{\alpha'}^{JM}(R) d\hat r_1 d\hat r_2 d\hat R,
\label{eq:cpld}
\end{eqnarray}
where $\alpha \equiv (j_1j_2j_{12}L)$.
We apply the hybrid modified log-derivative-Airy propagator in the general purpose scattering program MOLSCAT 
\cite{hutson94} to solve the coupled Eqs. (\ref{eq:cpld}). Additionally, we have tested other
propagator schemes included in MOLSCAT.
Our calculations revealed that other propagators can also produce quite stable results.

Boothroyd   {\it et al.} (BMKP) \cite{booth02} constructed 
a global six-dimensional PES for two hydrogen molecules,
especially  to represent the whole interaction region of the chemical reaction dynamics of the 
four-atomic system and to provide an accurate estimate of
the van der Waals well. The ground state 
and a few excited-state energies were calculated.
In the six-dimensional configuration space of the H$_2$-H$_2$ system the 
conical intersection forms a complicated three-dimensional hyper surface.
The new potential fits the van der Waals well to an accuracy of about 5\% \cite{booth02}.
In our calculation of the BMKP PES for H$_2$+H$_2$
the bond length was fixed at 1.449 a.u. or $r(\mbox{H}_2)$=0.7668 \r{A}
as in the Diep and Johnson (DJ) PES \cite{karl2000}. In the case of the HD+HD
calculation the  bond length of HD was adopted at $r(\mbox{HD})=0.7631$ \r{A}.

The log-derivative matrix of the wave function is propagated to large $R$-intermolecular 
distances, since all experimentally observable
quantum information about the collision is contained in the asymptotic behavior of functions 
$U^{JM}_{\alpha}(R\rightarrow\infty)$. The numerical results are matched to the 
known asymptotic solution to 
derive the physical scattering $S$-matrix
\begin{eqnarray}
U_{\alpha}^J
& \mathop{\mbox{\large$\sim$}}\limits_{R \rightarrow + \infty} &
\delta_{\alpha \alpha'}
e^{-i(k_{\alpha \alpha}R-(l\pi/2))} 
- \left(\frac{k_{\alpha \alpha}}{k_{\alpha \alpha'}}\right)^{1/2}S^J_{\alpha \alpha'}\nonumber \\
&\times &
e^{-i(k_{\alpha \alpha'}R-(l'\pi/2))},
\end{eqnarray}
where $k_{\alpha \alpha'}=[2M_{12}(E+E_{\alpha}-E_{\alpha'})]^{1/2}$ 
is the channel wave number, $E_{\alpha(\alpha')}$
are rotational channel energies and $E$ is the total energy in the $(1234)$ system.
The method was used for each partial wave until a converged cross section was obtained.

\begin{table*}
\caption{Rotational channel energies in the two hydrogen systems:
a). HD + HD and b). $para$-H$_2$ + $para$-H$_2$}
\vspace{3mm}
\centering
\label{table_1}
\begin{ruledtabular}
\begin{tabular}{cccclcccccl}\\
\multicolumn{5}{c}{a). HD($j_1$)$ + $HD($j_2$)}&  &\multicolumn{5}{c}{b). $para$-H$_2(j_1)$$ + para$-H$_2(j_2)$}\\ \\
\hline \\
$j_1$&$j_2$&$j_{12}$&$\nu_a$&$\epsilon^{HD}_{j_1 j_2}(\nu_a), $\ cm$^{-1}$ &  &
$j_1$&$j_2$&$j_{12}$&$\nu_b$&$\epsilon^{H_2}_{j_1 j_2}(\nu_b), $\ cm$^{-1}$\\ \\
\hline
0 &  0 &   0 & 1 &  0.0     & &   0  &     0 &      0 &      1 &            0.0\\
0 &  1 &   1 & 2 &  89.4    & &   0  &     2 &      2 &      2 &            364.8\\
0 &  2 &   2 & 3 &  268.2   & &   0  &     4 &      4 &      3 &            1216.0\\ \\
1 &  1 &   0 & 4 &  178.8   & &   2  &     2 &      0 &      4 &            729.6\\
1 &  1 &   1 & 4 &  178.8   & &   2  &     2 &      1 &      4 &            729.6\\
1 &  1 &   2 & 4 &  178.8   & &   2  &     2 &      2 &      4 &            729.6\\
  &    &     &   &          & &   2  &     2 &      3 &      4 &            729.6\\
  &    &     &   &          & &   2  &     2 &      4 &      4 &            729.6\\  \\
1 &  2 &   1 & 5 &  357.6   & &   2  &     4 &      2 &      5 &            1580.8\\
1 &  2 &   2 & 5 &  357.6   & &   2  &     4 &      3 &      5 &            1580.8\\
1 &  2 &   3 & 5 &  357.6   & &   2  &     4 &      4 &      5 &            1580.8\\
  &    &     &   &          & &   2  &     4 &      5 &      5 &            1580.8\\
  &    &     &   &          & &   2  &     4 &      6 &      5 &            1580.8\\ \\
2 &  2 &   0 & 6 &  536.4   & &   4  &     4 &      0 &      6 &            2432.0\\
2 &  2 &   1 & 6 &  536.4   & &   4  &     4 &      1 &      6 &            2432.0\\
2 &  2 &   2 & 6 &  536.4   & &   4  &     4 &      2 &      6 &            2432.0\\
2 &  2 &   3 & 6 &  536.4   & &   4  &     4 &      3 &      6 &            2432.0\\
2 &  2 &   4 & 6 &  536.4   & &   4  &     4 &      4 &      6 &            2432.0\\
  &    &     &   &          & &   4  &     4 &      5 &      6 &            2432.0\\
  &    &     &   &          & &   4  &     4 &      6 &      6 &            2432.0\\
  &    &     &   &          & &   4  &     4 &      7 &      6 &            2432.0\\
  &    &     &   &          & &   4  &     4 &      8 &      6 &            2432.0\\
\end{tabular}
\end{ruledtabular}
\end{table*}

\begin{table*}
\caption{Comparison between different but ``corresponding" $(\nu_a=\nu_b)$
state-resolved cross-sections (\AA$^2$) in the HD + HD and
$para$-H$_2$ + $para$-H$_2$ collisions at ultracold $T=1.439 \cdot 10^{-8}$ K and very high $T=14390.0$
K temperatures.} 
\vspace{3mm}
\centering
\label{table_2}
\begin{ruledtabular}
\begin{tabular}{clcccclclccccl}\\
\multicolumn{7}{c}{HD($j_1$)+HD($j_2$) $\rightarrow$ HD($j'_1$) + HD($j'_2$)}&  
\multicolumn{7}{c}{H$_2(j_1)$ + H$_2(j_2)$ $\rightarrow$ H$_2(j'_1)$ + H$_2(j'_2)$}\\ \\
\hline \\
$E_{kin}$, K & $\epsilon^{HD}_{j_1j_2}(\nu)$ & $j_1$&$j_2$&$j'_1$&$j'_2$&
$\sigma^{HD}_{j_1j_2\rightarrow j'_1j'_2}$
&  & 
$\epsilon^{H_2}_{j_1j_2}(\nu)$& $j_1$&$j_2$&$j'_1$&$j'_2$&
$\sigma^{H_2}_{j_1j_2\rightarrow j'_1j'_2}$\\ \\
\hline
1.439$\times$10$^{-8}$&89.4&  0 &  1 &   0 & 0 &  1.00$\times10^5$    &     &  364.8  &   0  &  2 &  0 &  0 &  0.65$\times 10^2$\\
                                       &      &  0 &  0 &   0 & 1 &  3.34$\times10^{-5}$&     &             &   0  &  0 &  0 &  2 &  0.89$\times 10^{-8}$\\
                                  & 178.8&  1 &  1 &   0 & 1 &  1.94$\times10^4$     &     &  729.6   &   2  &  2 &  0 &  2 &  2.06$\times 10^2$\\
                                  &          &  1 &  1 &   0 & 0 &  0.50$\times10^4$     &     &              &   2  &  2 &  0 &  0 &  17.8$                  $\\
%
%
                                  & 536.4&  2 &  2 &   1 & 1 &  0.52$\times10^4$     &     &  2432.0 &   4  &  4 &  2 &  2 &  2.31$                      $\\
                                  &          &  2 &  2 &   0 & 2 &  0.55$\times10^3$     &     &              &   4  &  4 &  0 &  4 &  1.12$                      $\\
                                  &          &  2 &  2 &   0 & 1 &  0.94$\times10^3$     &     &              &   4  &  4 &  0 &  2 &  0.50$\times10^{-1}$\\
                                  &          &  2 &  2 &   0 & 0 &  1.28$\times10^2$     &     &              &   4  &  4 &  0 &  0 &  1.94$\times10^{-3}$\\
\vspace{4mm}\\
1.439$\times$10$^{4}$& 89.4&  0 &  1 &   0 & 0&  0.60$                $      &      &  364.8 &   0  &  2 &  0 &  0  &  0.25$                     $\\
                                  &         &  0 &  0 &   0 & 1 &  1.78 $                $      &      &            &   0  &  0 &  0 &  2  &  1.18$                      $\\
                                  & 357.6&  1 &  2 &   1 & 1 &  1.07$                $      &     &  1580.8 &   2  &  4 &  2 &  2 &  0.44\\ 
                                  &          &  1 &  2 &   0 & 2 &  0.57$                $      &     &              &   2  &  4 &  0 &  4 &  0.374\\ 
\end{tabular}
\end{ruledtabular}
\end{table*}

Cross sections for rotational excitation and relaxation   can be obtained directly from the $S$-matrix.
In particular the cross sections for excitation from $j_1j_2\rightarrow j'_1j'_2$ summed over the final $m'_1m'_2$
and averaged over the initial $m_1m_2$ are given by
\begin{eqnarray}
\sigma(j'_1,j'_2;j_1j_2,\varepsilon)=\sum_{Jj_{12}j'_{12}LL'}
\frac{\pi(2J+1)}{(2j_1+1)(2j_2+1)k_{\alpha\alpha'}}\nonumber \\
\times |\delta_{\alpha\alpha'}- 
S^J(j'_1,j'_2,j'_{12}L';j_1,j_2,j_{12},L; E)|^2.
\label{eq:cross}
\end{eqnarray}
The kinetic energy is
$\varepsilon=E-B_1j_1(j_1+1)-B_2j_2(j_2+1)$,
where $B_{1(2)}$ are the rotation constants of rigid rotors $(12)$ and $(34)$ respectively.

The relationship between the rate coefficient $k_{j_1j_2\rightarrow j'_1j'_2}(T)$ and the corresponding
cross section 
$\sigma_{j_1j_2\rightarrow j'_1j'_2}(\varepsilon)$
can be obtained through the following weighted average
\begin{eqnarray}
k_{j_1j_2\rightarrow j'_1j'_2}(T) & = &
\sqrt{\frac{8k_BT}{\pi\mu}}
\frac{1}{(k_BT)^2}\int_{\varepsilon_s}^{\infty}
\sigma_{j_1j_2\rightarrow j'_1j'_2}(\varepsilon)\nonumber \\
&\times &
e^{-\varepsilon/k_BT}\varepsilon d\varepsilon,
\label{eq:rate}
\end{eqnarray}
where 
$k_B$ is Boltzmann constant, $\mu$ is reduced mass of the 
molecule-molecule system and $\varepsilon_s$ is the minimum kinetic energy for the levels $j_1$ and $j_2$
to become accessible.

\section{Results}
In this section our numerical results for rotational transitions in  
HD+HD  collision
and $para/para$-hydrogen molecules are presented. We carry out state-to-state
comparison between these two collisions for selected rotational transitions in the HD and H$_2$ molecules.
Specifically the following rotational energy transfer processes are considered:
\begin{eqnarray}
\mbox{HD}(j_1) +\mbox{HD}(j_2) & \rightarrow & \mbox{HD}(j'_1) + \mbox{HD}(j'_2),
\label{eq:hdhd} \\
\mbox{H}_2(j_1) +\mbox{H}_2(j_2) &  \rightarrow & \mbox{H}_2(j'_1) + \mbox{H}_2(j'_2).
\label{eq:h2h2} \end{eqnarray}

At first look one might expect that the scattering outputs of these two collisions 
(\ref{eq:hdhd}) and (\ref{eq:h2h2}) should be close to each other. This is because
the PESs of H$_2$-H$_2$ and HD-HD are almost the same six-dimensional functions of the
H$_4$ four-atomic system coordinates. This fact follows from the general idea of
the Born-Oppenheimer model \cite{BO} and simple theoretical atom-molecular
consideration. Therefore, the two processes (\ref{eq:hdhd}) and (\ref{eq:h2h2}) should 
lead to similar results.
At the same time the HD and H$_2$ molecules
have different rotational constants. This difference is not dramatic: 
the rotational constant of H$_2$ is $B_e(\mbox{H}_2)=60.8$ cm$^{-1}$, but the same parameter
for  HD is $B_e(\mbox{HD})=44.7$ cm$^{-1}$.  
  

The HD+HD system has only four electrons. Further, the 
HD molecule consists of two nonidentical atoms which are in a covalent bond. In 
covalent bonding the spins of the electrons are antiparallel. The interaction of one 
of the nuclei, H$^+$ or D$^+$, with its own electron leads to a quantum 
configuration where the spin of its electron is oriented antiparallel to the spin of 
the nucleus, i.e. H$^+$ or D$^+$. Thus, the spins of H$^+$ and D$^+$ are 
antiparallel. Because the spin of H$^+$ is $I_1=-1/2$ and the spin of D$^+$ is 
$I_2=1$ the resulting spin of the HD molecule nucleus is $I_{12}=1/2$. This value 
has been adopted in the current calculation, although there may be other possible 
values. The processes (\ref{eq:hdhd}) and (\ref{eq:h2h2}) are collisions between two 
indistinguishable diatomic molecules. This fact is taken into account in this computation.


\subsection{Comparison between HD+HD and H$_2$+H$_2$ state-selected 
integral cross sections} The precise HD-HD PES can be derived from the 
H$_2$-H$_2$ surface by adjusting, i.e. shifting, the coordinates of the 
center of masses of the two H$_2$ molecules to the center of masses of 
the HD molecules. Once the symmetry is broken in H$_2$-H$_2$ by 
replacing the H atoms with the D atoms in the two H$_2$ molecules we obtain 
the full HD-HD PES. The new potential will posses all parts of the HD-HD 
interaction. Therefore, it will be interesting to consider scatteing in 
two systems, which are not very different like HD+HD and H$_2$+H$_2$.

In this work a large number of test calculations have been done to 
secure the convergence of the results with respect to all parameters 
that enter into the propagation of the Schr\"odinger equation 
(\ref{eq:schred}). This includes the intermolecular distance $R$, the 
total angular momentum $J$ of the four atomic system, the number of 
rotational levels $N_{lvl}$ to be included in the close coupling 
expansion and some others, see the MOLSCAT manual \cite{hutson94}. We 
reached convergence for the integral cross sections, $\sigma(E_{kin})$, 
in all cases. However, it was particularly difficult to achieve 
convergence on the parameter $R$ in both cases. For the applied BMKP PES 
we used $R_{min}=2$ \r{A} to $R_{max}=50$ \r{A}. We also applied a few 
different propagators included in the MOLSCAT program.

\begin{figure}[!t]
\begin{center}
\includegraphics*[scale=1.0,width=20pc,height=12pc]{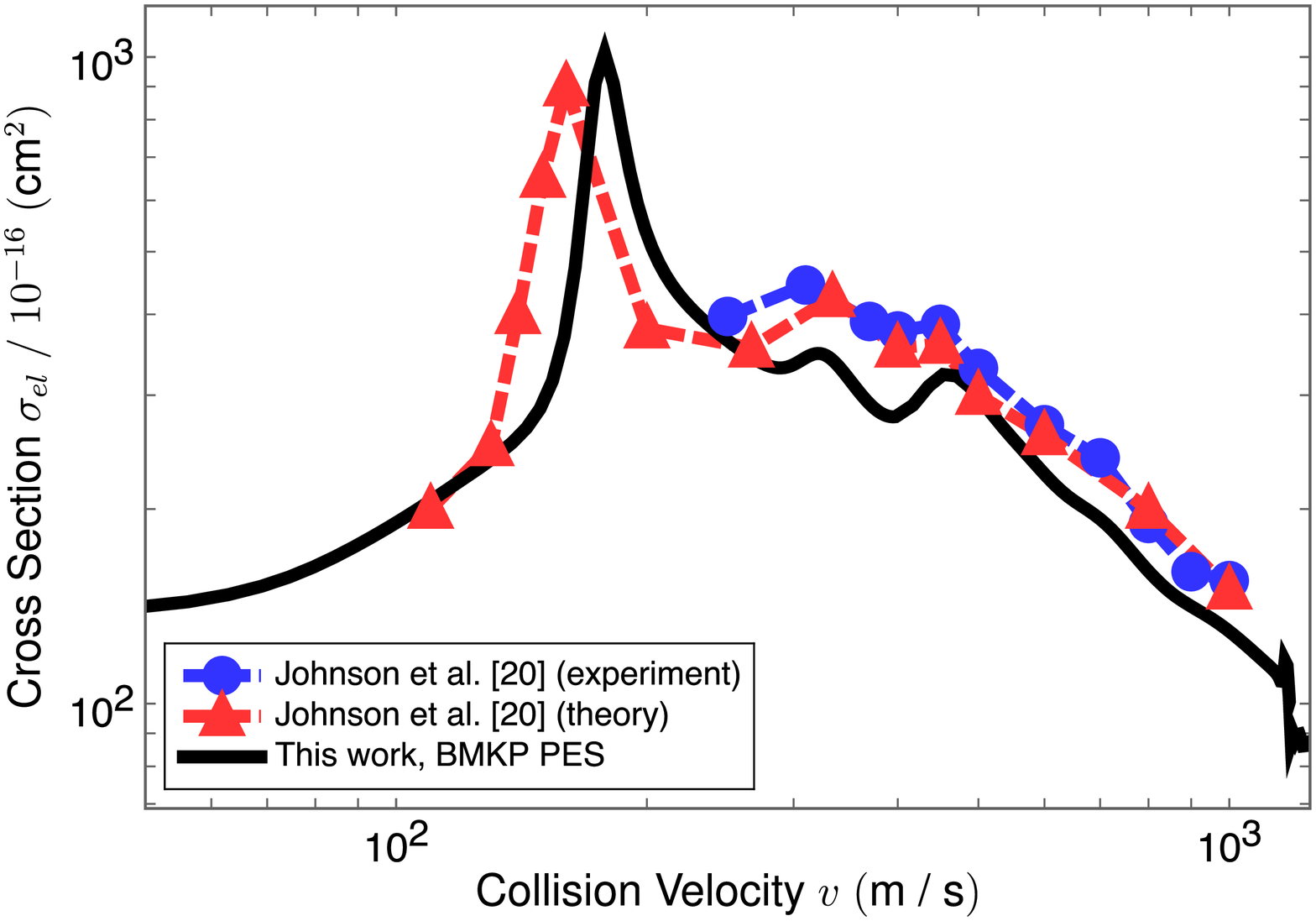}  
\vspace{1mm}\\
%
\end{center} 

\caption{(Color online)   Integral cross sections for the
HD+HD elastic scattering computed with the modified BMKP PES \cite{booth02}
together with the experimental and theoretical data from work \cite{johnson79}.
 }
\label{fig2} \end{figure}

In a previous paper we presented a detailed description of convergence 
test for H$_2$+H$_2$ collision \cite{renat06}. The same numerical 
convergence has been achieved in this work. Namely, stable total cross 
sections have been obtained with respect to the number $N_{lvl}$ of the 
rotational levels to be included in the basis set (\ref{eq:expn}) of 
HD+HD, i. e.,  in each HD molecule the discrete integer quantum numbers 
$j_1$ and $j_2$ run from 0 to 4. As a result a maximum number $N_{lvl} 
(=55)$ of rotational levels in HD+HD has been generated. With regard to 
the total quantum angular momentum $J$ in the HD+HD system at $T\sim 
10^{-8}-10^{-6}$ K we needed to adopt just a few discrete values of this 
parameter, that is $J_{max}=$3 or 4 was quite enough. However, at larger 
collision energies $E_{kin}$, such as, the ones we considered at 
$T\sim4000$ K, one needs up to $J_{max}\approx$60. In the case of 
$E_{kin}\sim14000$ K: $J_{max}\approx$ 120.

We present in Table I the rotational channel energies in the 
HD$-$HD and para-H$_2$-para-H$_2$ systems. This is a comparative 
table of the rotational spectra of these two systems. The 
first five columns from the left present HD$-$HD and the other 
five columns present the para-H$_2$-para-H$_2$ system; $j_1$ and 
$j_2$ are the quantum orbital momenta of the HD and H$_2$ 
molecules, $\vec j_{12}=\vec j_1+\vec j_2$, with $|j_1-j_2| \le 
j_{12} \le j_1+j_2$, the index $\nu_{a(b)}$ is the current number 
of the degenerate rotational levels in HD and H$_2$, respectively. 
The  rotational energy levels are shown in cm$^{-1}$. 
The goal of this work is to investigate the ultracold regime in 
HD+HD, calculate its rotational energy transfer cross sections and 
thermal rate coefficients, and to carry out comparison with the 
corresponding (when $\nu_a=\nu_b$, see Table I) rotational 
transitions in   
para-H$_2$-para-H$_2$ collision.  


There are two slightly different definitions \cite{danby87} of 
the rotational cross sections in  collisions between 
two identical diatomic molecules, for example, in 
HD($j_1$)+HD($j_2$)$\rightarrow$ HD($j'_1$)+HD($j'_2$).  The cross 
section for the rotational transition $j_1 j_2\rightarrow j'_1 
j'_2$ is \cite{green75,hutson94}: $\sigma \sim (1 + \delta_{j_1 
j_2})(1 + \delta_{j'_1 j'_2})$. However, in Ref. \cite{monchick80} 
the same cross section has been defined as $\sigma \sim (1 + 
\delta_{j_1 j_2} \delta_{j'_1 j'_2})$. It is seen that the two 
cross sections coincide when $j_1 \neq j_2$ and $j'_1 \neq j'_2$. 
However, for other combinations of the rotational quantum numbers, 
namely, when $j_1 = j_2$ and/or $j'_1 = j'_2$,  the cross section 
calculated in accord with Refs. \cite{green75,hutson94} is two 
times larger than the cross section from Ref. \cite{monchick80}. 
This has been taken into account in  calculation with the 
MOLSCAT program \cite{hutson94}, i.e.,  for the integral cross 
sections $\sigma(j'_1 j'_2;j_1 j_2,\epsilon)$ from Eq. 
(\ref{eq:cross}) the following pre-factor $[(1 + \delta_{j_1 
j_2})(1 + \delta_{j'_1 j'_2})]^{-1}$ has been adopted.

First, let us turn to HD+HD elastic scattering. In Fig. \ref{fig2} 
we show results of this work computed with the modified BMKP PES 
\cite{booth02} together with the corresponding data 
(experiment/theory) from relatively old papers \cite{johnson79}. 
As can be seen all these cross sections are in a satisfcatory 
agreement with each other. This test calculation reveals the 
reliability of the modified BMKP PES, the computer program and the 
numerical convergence. 
One can see that at low and very low 
energies the general forms of the cross sections are rather close 
to each other with the exception of a shape resonance 
and small oscillations in 
the cross sections in Fig. 2. 

\begin{figure}
\begin{center}
\includegraphics*[scale=1.0,width=20pc,height=12pc]{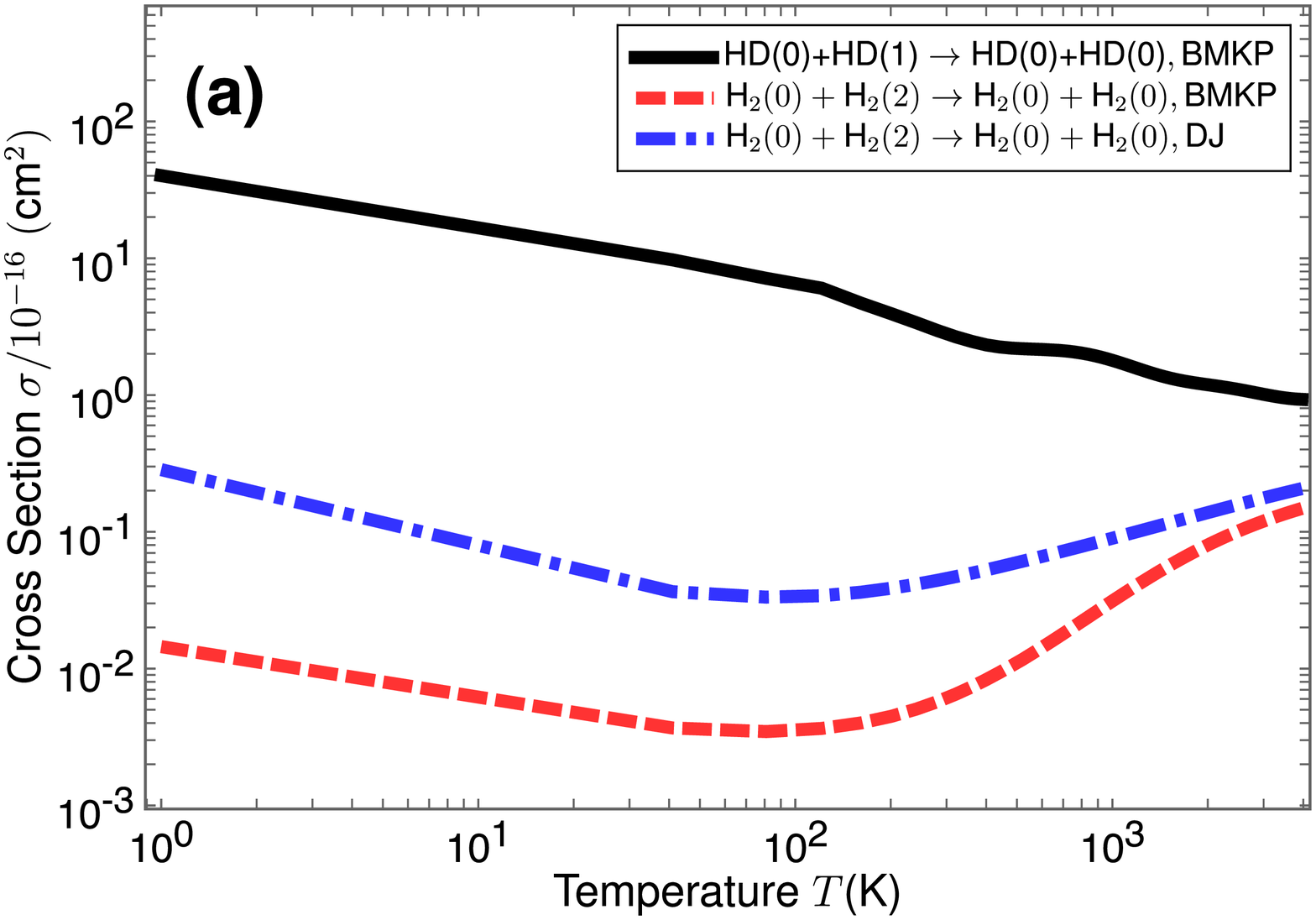}   
\vspace{1mm}\\
\includegraphics*[scale=1.0,width=20pc,height=12pc]{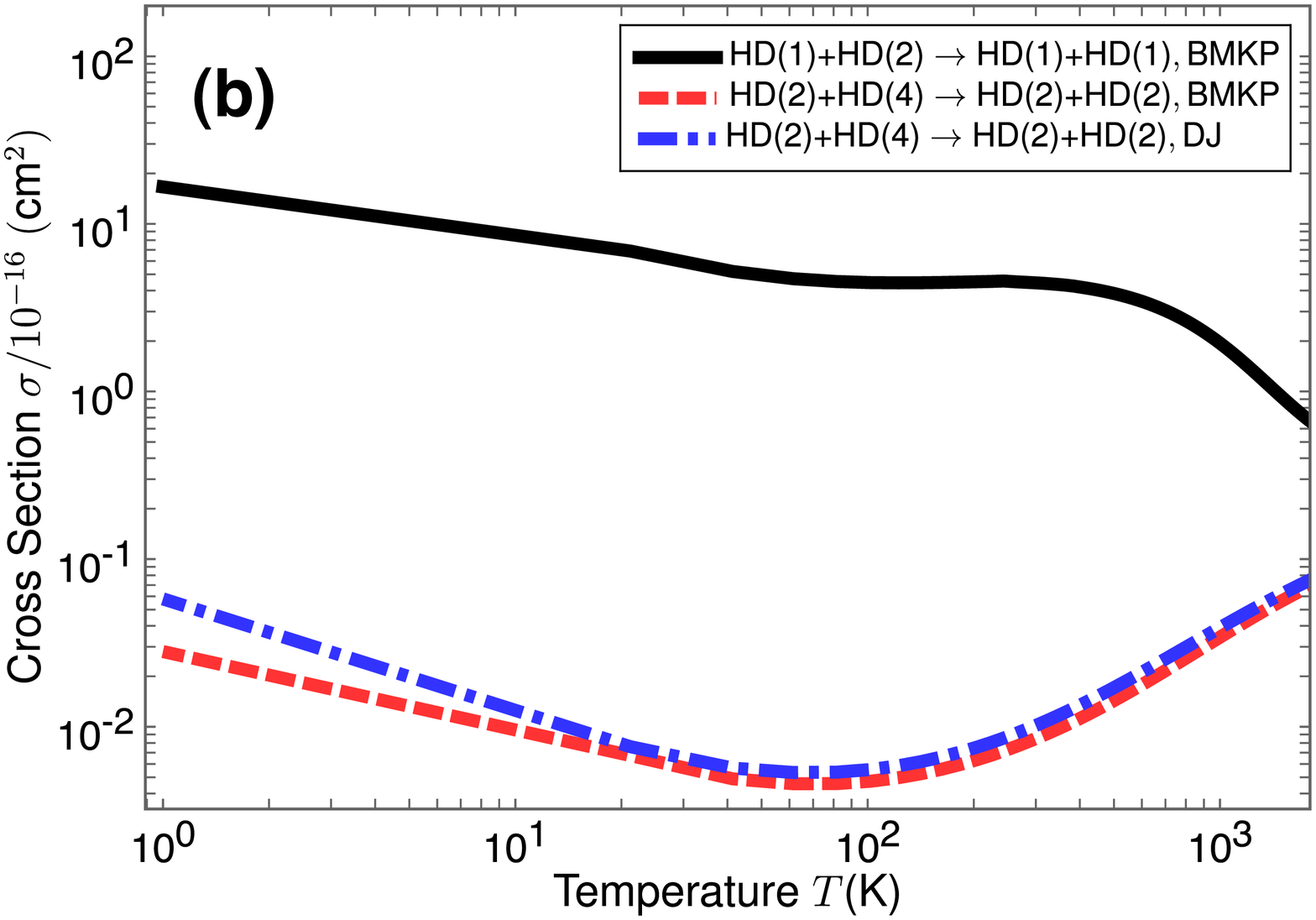}  
%

\caption{(Color online) (a) Inelastic scattering integral cross 
sections for $\mbox{HD}(0) +\mbox{HD}(1) \rightarrow \mbox{HD}(0) 
+ \mbox{HD}(0)$ and $\mbox{H}_2(0) +\mbox{H}_2(2) \rightarrow 
\mbox{H}_2(0) + \mbox{H}_2(0)$.  (b) The same for  $\mbox{HD}(1) 
+\mbox{HD}(2) \rightarrow \mbox{HD}(1) + \mbox{HD}(1)$ and 
$\mbox{H}_2(2) +\mbox{H}_2(4) \rightarrow \mbox{H}_2(2) + 
\mbox{H}_2(2)$.} \label{fig3} \end{center} \end{figure}

\begin{figure}
\begin{center}
\includegraphics*[scale=1.0,width=20pc,height=12pc]{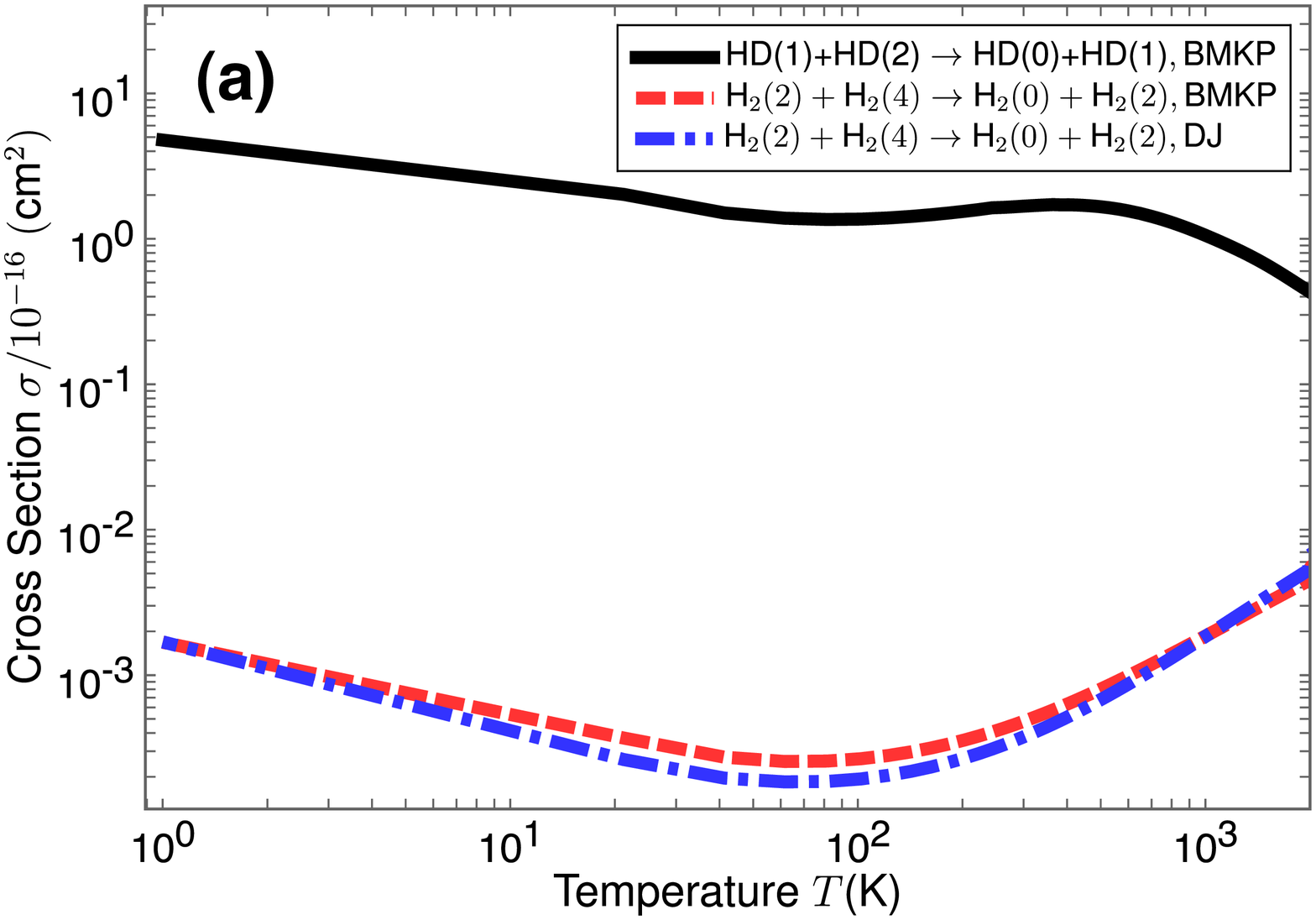}  
\vspace{1mm}\\
\includegraphics*[scale=1.0,width=20pc,height=12pc]{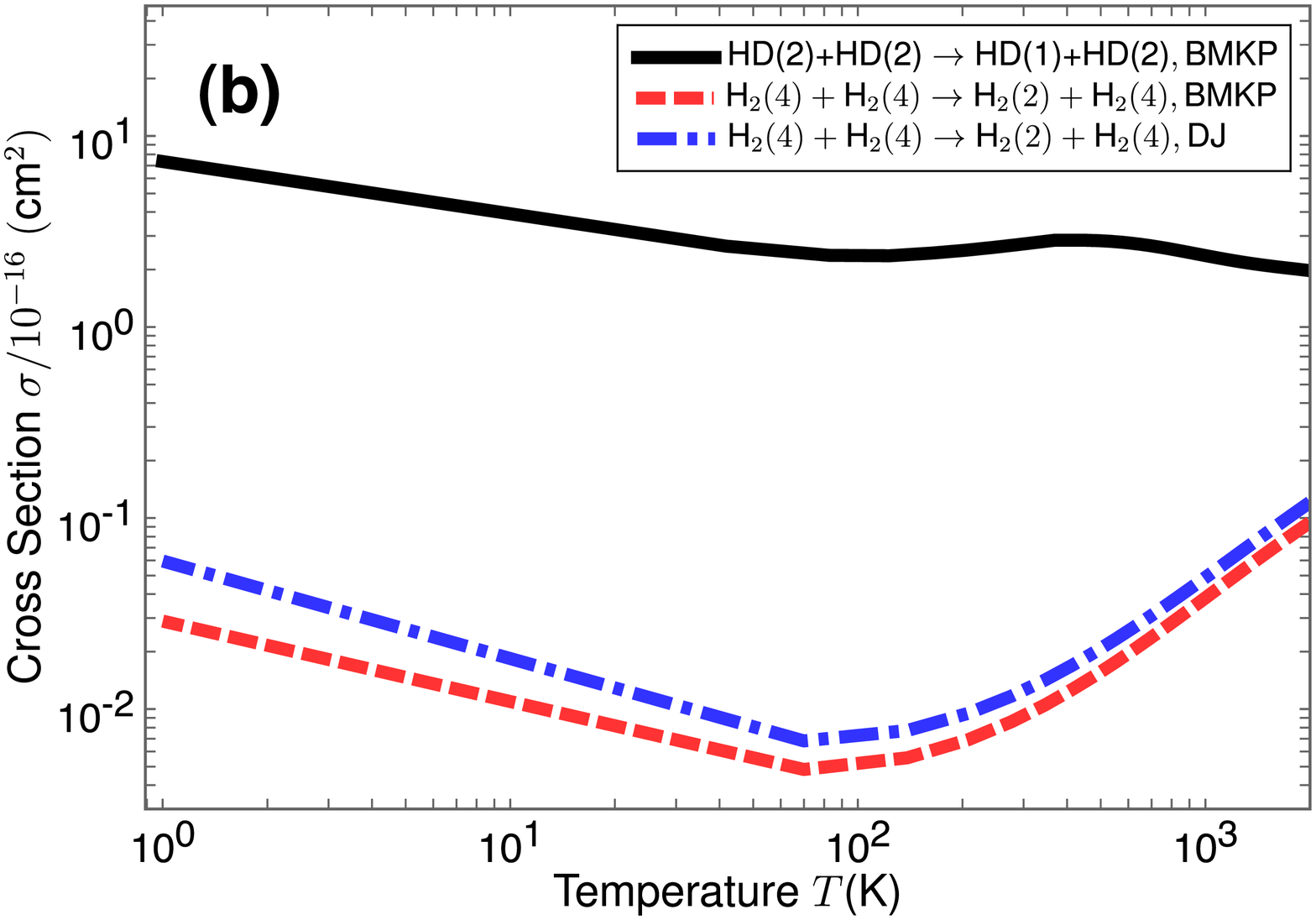}  
\caption{(Color online)  (a)
Inelastic scattering integral cross sections for
$\mbox{HD}(1) +\mbox{HD}(2) \rightarrow \mbox{HD}(0) + \mbox{HD}(1)$ and
$\mbox{H}_2(2) +\mbox{H}_2(4) \rightarrow \mbox{H}_2(0) + \mbox{H}_2(2)$.
 (b) The same for 
$\mbox{HD}(2) +\mbox{HD}(2) \rightarrow \mbox{HD}(1) + \mbox{HD}(2)$ and
$\mbox{H}_2(4) +\mbox{H}_2(4) \rightarrow \mbox{H}_2(2) + \mbox{H}_2(4)$.}
\label{fig4} \end{center} \end{figure}

\begin{figure}
\begin{center}
\includegraphics*[scale=1.0,width=20pc,height=12pc]{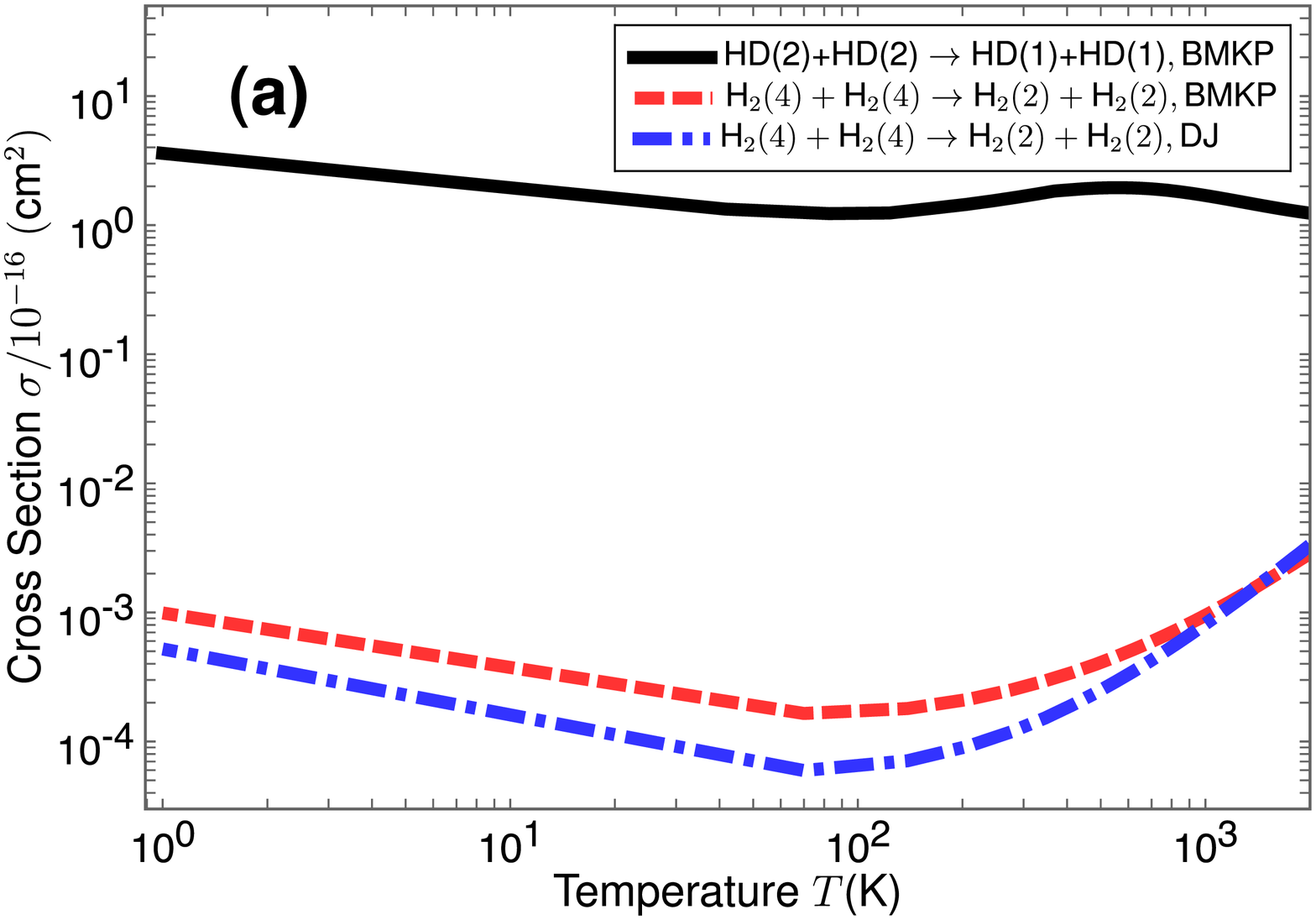} 
\vspace{1mm}\\
\includegraphics*[scale=1.0,width=20pc,height=12pc]{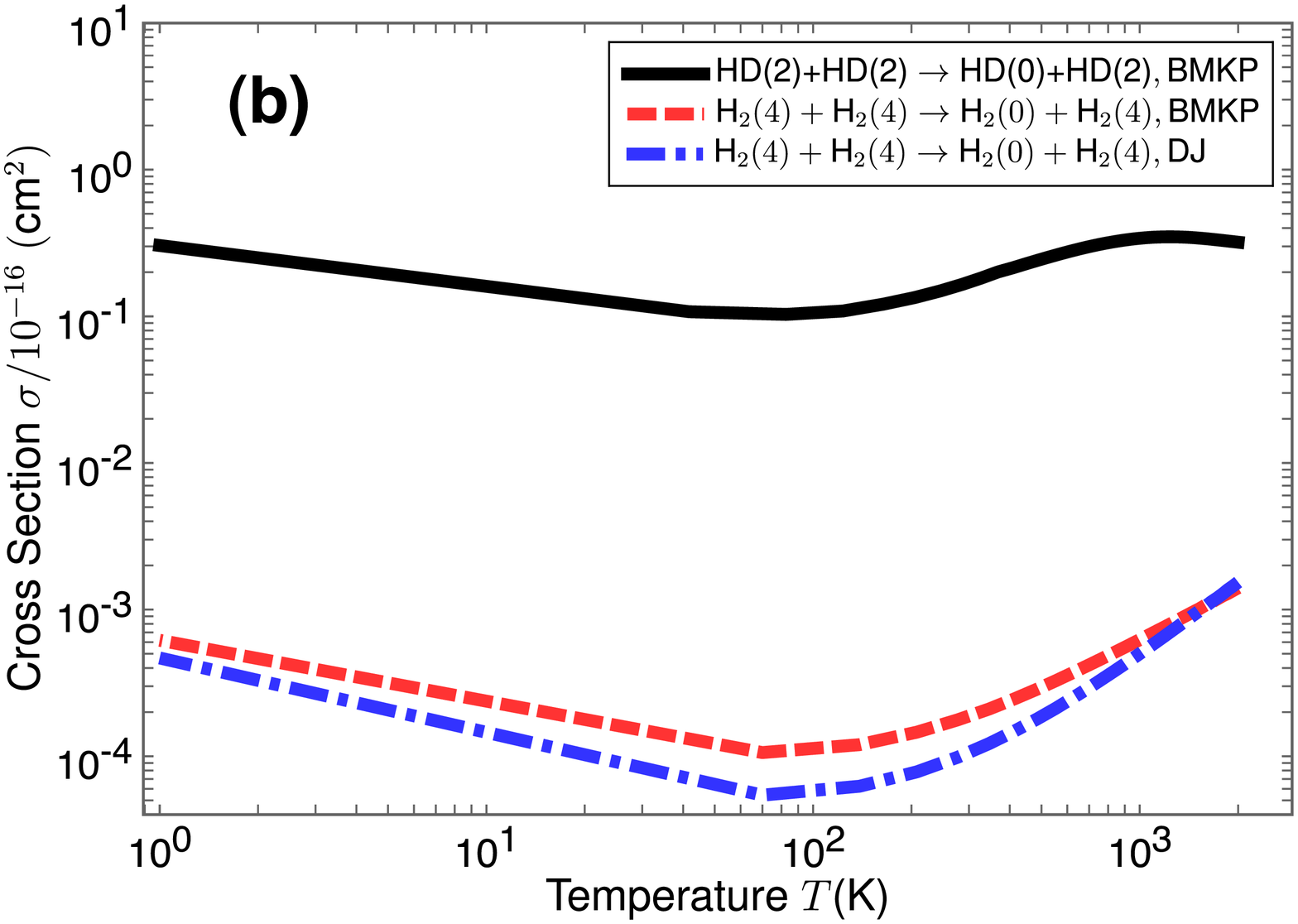} 
\caption{(Color online) (a)
Inelastic scattering integral cross sections for
$\mbox{HD}(2) +\mbox{HD}(2) \rightarrow \mbox{HD}(1) + \mbox{HD}(1)$ and
$\mbox{H}_2(4) +\mbox{H}_2(4) \rightarrow \mbox{H}_2(2) + \mbox{H}_2(2)$.
 (b) The same for 
$\mbox{HD}(2) +\mbox{HD}(2) \rightarrow \mbox{HD}(0) + \mbox{HD}(2)$ and
$\mbox{H}_2(4) +\mbox{H}_2(4) \rightarrow \mbox{H}_2(0) + \mbox{H}_2(4)$.}
\label{fig5} \end{center} \end{figure}


In Table II few selected state-to-state cross sections for the HD+HD and H$_2$+H$_2$
collisions are presented. Here we compare results for few specific rotational excitation and de-excitation
integral cross sections at only two values of kinetic energy,
namely at  $T_1\sim 10^{-8}$ K, and at very high collision energy, i.e. $T _2\sim 14,000$ K.
 At low temperature the HD+HD cross section could be larger by three to four orders of magnitude, whereas 
at high temperature the two cross sections are of the same order.  

In Figs. \ref{fig3}, \ref{fig4} and \ref{fig5} we present a few 
state-selected rotational transition cross sections in collisions 
(\ref{eq:hdhd}) and (\ref{eq:h2h2}). It is useful to see the 
corresponding cross sections together on a single plot, that is 
when $\nu_a = \nu_b$. For example, in Fig. \ref{fig3} (a) we show 
rotational transition de-excitation
cross sections from the first excited states 
of HD and H$_2$ molecules, i.e. we consider $\mbox{HD}(0) 
+\mbox{HD}(1) \rightarrow \mbox{HD}(0) + \mbox{HD}(0)$ and 
$\mbox{H}_2(0) +\mbox{H}_2(2) \rightarrow \mbox{H}_2(0) + 
\mbox{H}_2(0)$ for a wide range of kinetic energies: from 1 K to 
up to 4000 K. In the case of H$_2$+H$_2$, we carry out 
computations with two different PESs, e.g., with the BMKP PES 
\cite{booth02} and with the Diep-Johnson (DJ) H$_2$-H$_2$ PES from 
Ref. \cite{karl2000}. The last one was formulated for fixed 
equilibrium distances between the hydrogen atoms in each H$_2$ 
molecule. In Fig. \ref{fig3} (b) we show cross sections for 
some other de-excitation processes in the HD-HD system. 



Further results for the integral cross section are shown in Figs. 
\ref{fig4} and \ref{fig5}. The cross sections 
for following processes are presented in Fig. \ref{fig4} (a): 
$\mbox{HD}(1) +\mbox{HD}(2) \rightarrow \mbox{HD}(0) + 
\mbox{HD}(1)$ and $\mbox{H}_2(2) +\mbox{H}_2(4) \rightarrow 
\mbox{H}_2(0) + \mbox{H}_2(2)$. The cross sections of 
$\mbox{HD}(2) +\mbox{HD}(2) \rightarrow \mbox{HD}(1) + 
\mbox{HD}(2)$ and $\mbox{H}_2(4) +\mbox{H}_2(4) \rightarrow 
\mbox{H}_2(2) + \mbox{H}_2(4)$ are shown in Fig. \ref{fig4} (b). 
It is seen in Fig. \ref{fig4} (a) that the HD+HD de-excitation 
cross sections could be larger than the H$_2$+H$_2$ de-excitation
cross sections by four orders of magnitudes.
For the cross sections in  Fig. \ref{fig4} (b) the difference is 
about a factor 
of 10$^3$. In Fig. \ref{fig5} (a) we show results for rotational 
transitions: $\mbox{HD}(2) +\mbox{HD}(2) \rightarrow \mbox{HD}(1) 
+ \mbox{HD}(1)$ and $\mbox{H}_2(4) +\mbox{H}_2(4) \rightarrow 
\mbox{H}_2(2) + \mbox{H}_2(2)$. In Fig. \ref{fig5} (b), we show 
the same for the following transitions. $\mbox{HD}(2) 
+\mbox{HD}(2) \rightarrow \mbox{HD}(0) + \mbox{HD}(2)$ and 
$\mbox{H}_2(4) +\mbox{H}_2(4) \rightarrow \mbox{H}_2(0) + 
\mbox{H}_2(4)$. Again, in both cases the  HD+HD cross sections could 
be larger than the H$_2$+H$_2$ at 1 K by four orders of magnitudes.
We  see in Figs. 3, 4, and 5 that,  for the  
de-excitation processes for H$_2$+H$_2$, 
the BMKP and the DJ PESs provide similar results with the exceptions 
shown in Fig.\ 2.

In Ref.  \cite{renat06a} it was shown that for a specific excitation
rotational transition in the H$_2$+H$_2$ inelastic scattering, e.g., in  
$
\mbox{H}_2(0) +\mbox{H}_2(0) \rightarrow \mbox{H}_2(0) + \mbox{H}_2(2),
$
the BMKP PES provides an incorrect cross section when compared to
the DJ potential. The comparison was also  carried out with
available experimental data \cite{mate2005}.
Nevertheless, the BMKP PES has been applied to the important (astrophysical) 
$o$-/$p$-H$_2$+HD inelastic scattering problem \cite{renat07,renat09}.
This is why we applied the BMKP PES to the HD+HD scattering problem.


In conclusion, based on the Born-Oppenheimer model treatment, the 
HD+HD and H$_2$+H$_2$ have similar
PESs. However, in the case of the HD-HD system the original 
H$_2$-H$_2$ PES is adopted and
the two center of masses of both H$_2$ molecules is just slightly 
shifted to the appropriate
positions of the HD molecule center of masses.
After this procedure we obtain the full space, i.e. global HD-HD PES.  
Our computations with this modified PES revealed a very strong isotopic
effect in the HD+HD and H$_2$+H$_2$ collisions at low energies.

\begin{figure}[!t]
\begin{center}
\includegraphics*[scale=1.0,width=20pc,height=12pc]{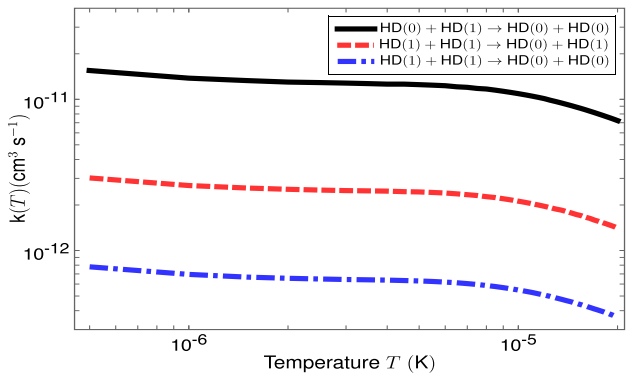} 
\caption{(Color online) 
Thermal rate coefficients for the inelastic scattering processes:
$\mbox{HD}(0)+\mbox{HD}(1)\rightarrow \mbox{HD}(0)+\mbox{HD}(0)$,
$\mbox{HD}(1)+\mbox{HD}(1)\rightarrow \mbox{HD}(0)+\mbox{HD}(1)$, and
$\mbox{HD}(1)+\mbox{HD}(1)\rightarrow \mbox{HD}(0)+\mbox{HD}(0)$,
at ultracold temperatures.}
\label{fig6} \end{center} \end{figure}

\begin{figure}[!b]
\begin{center}
\includegraphics*[scale=1.0,width=20pc,height=12pc]{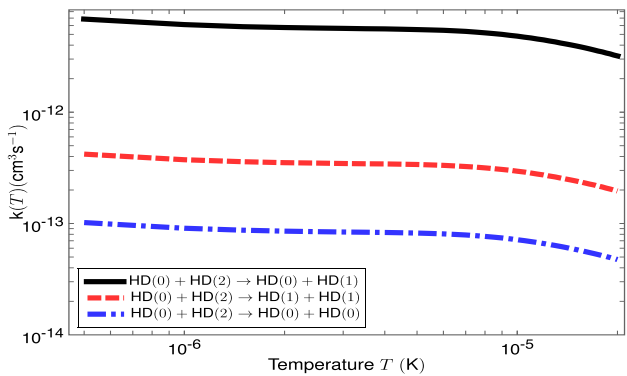}  
\caption{(Color online) The same as in Fig. \ref{fig6} for following processes:
$\mbox{HD}(0)+\mbox{HD}(2)\rightarrow \mbox{HD}(0)+\mbox{HD}(1)$,
$\mbox{HD}(0)+\mbox{HD}(2)\rightarrow \mbox{HD}(1)+\mbox{HD}(1)$, and
$\mbox{HD}(0)+\mbox{HD}(2)\rightarrow \mbox{HD}(0)+\mbox{HD}(0)$.}
\label{fig7} \end{center} \end{figure}

\subsection{HD+HD rotational state-selected thermal rate 
coefficients at ultracold temperatures}
\begin{figure}[!t]
\begin{center} 
\includegraphics*[scale=1.0,width=20pc,height=12pc]{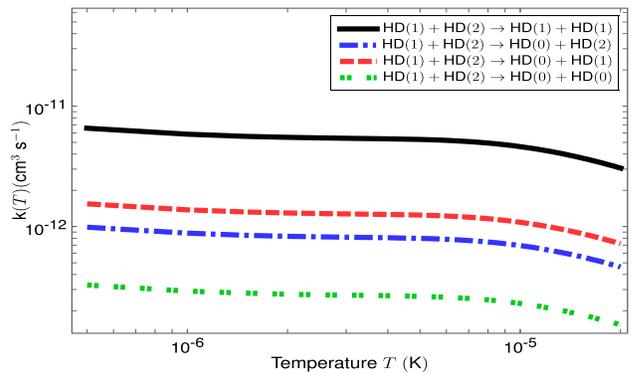}   
\caption{(Color online) 
 The same as in Fig. \ref{fig6} for following processes:
$\mbox{HD}(1)+\mbox{HD}(2)\rightarrow \mbox{HD}(1)+\mbox{HD}(1)$,
$\mbox{HD}(1)+\mbox{HD}(2)\rightarrow \mbox{HD}(0)+\mbox{HD}(2)$,
$\mbox{HD}(1)+\mbox{HD}(2)\rightarrow \mbox{HD}(0)+\mbox{HD}(1)$, and
$\mbox{HD}(1)+\mbox{HD}(2)\rightarrow \mbox{HD}(0)+\mbox{HD}(0)$.}
\label{fig8} \end{center} \end{figure}

\begin{figure}
\begin{center} 
\includegraphics*[scale=1.0,width=20pc,height=12pc]{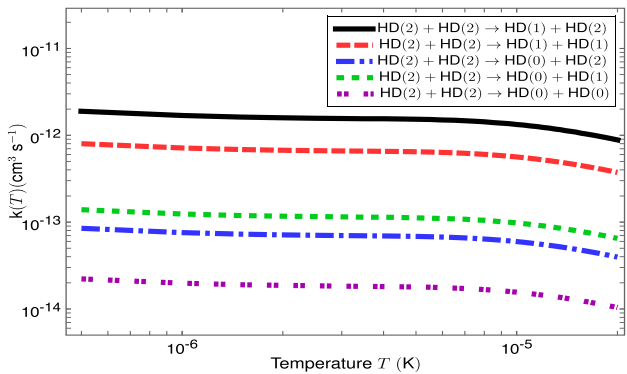}  
\caption{(Color online)
 The same as in Fig. \ref{fig6} for following processes:
$\mbox{HD}(2)+\mbox{HD}(2)\rightarrow \mbox{HD}(1)+\mbox{HD}(2)$,
$\mbox{HD}(2)+\mbox{HD}(2)\rightarrow \mbox{HD}(1)+\mbox{HD}(1)$,
$\mbox{HD}(2)+\mbox{HD}(2)\rightarrow \mbox{HD}(0)+\mbox{HD}(2)$,
$\mbox{HD}(2)+\mbox{HD}(2)\rightarrow \mbox{HD}(0)+\mbox{HD}(1)$, and
$\mbox{HD}(2)+\mbox{HD}(2)\rightarrow \mbox{HD}(0)+\mbox{HD}(0)$.}
\label{fig9} \end{center} \end{figure}

We show in Figs. \ref{fig6}, \ref{fig7}, \ref{fig8}, and \ref{fig9}    
the  thermal rate coefficients in the inelastic HD+HD collision
at very low temperatures from $\sim 5 \times 10^{-7}$ K to $\sim 2 \times 10^{-5}$ K.
These results were obtained from corresponding state-resolved integral 
cross sections 
$\sigma_{j_1j_2\rightarrow j_1'j_2}(\epsilon)$ with the use of 
expression (\ref{eq:rate}).
Only de-excitation thermal rates have been computed, because at such a low 
temperature 
 the excitation thermal rates are extremely small.
The rates have been computed for different initial rotational states of the 
HD molecules. 
The figure captions include the information about the specific
state-selected rotational transitions in both HD molecules before and after 
collision.

In Fig. \ref{fig6}, we show results for the rates       
$k_{j_1j_2\rightarrow j_1'j_2}(T)$ for the inelastic scattering processes
from the first three rotational states.
In Fig. \ref{fig7} we show the rates for transition  from the excited 
states. 
These transitions correspond to different de-excitations from lower 
initial rotational
states in the HD$-$HD system. 
In Fig. \ref{fig6} the initial state for the solid line corresponds to 
the first excited rotational state in HD+HD. From  Table I one 
can see that for this state
of the system: $\nu_a$=2. 
The two other lines in Fig. \ref{fig6}  
correspond to the initial state with $\nu_a$=4, 
i.e. $E_{in}$=178.8 cm$^{-1}$. The rates for the initial state 
HD(1)+HD(1)  in Fig. \ref{fig6} are smaller than those for the initial 
state HD(0)+HD(1), although all three rates have similar dependence on
temperature.  

In  Fig. \ref{fig7} we show the rates for  some de-excitation transitions
from
the initial state HD(0)+HD(2), which 
is the third rotational excited state in 
HD+HD: $\nu_a$=3. 
Although we again observe that the behavior 
of the thermal rates $k_{j_1j_2\rightarrow j_1'j_2}(T)$ is quite 
identical, their values significantly differ from each other, 
specifically up to two orders of magnitude. Further, the thermal rate 
coefficients $k_{j_1j_2\rightarrow j_1'j_2}(T)$ from the higher excited 
rotational states  HD(1)+HD(2) and HD(2)+HD(2) are presented in Fig. \ref{fig8} and 
\ref{fig9}, respectively.  
Specifically, 
these rates are from the energy levels 357.6 cm$^{-1}$ and 536.4 
cm$^{-1}$ corresponding to the following two 
indices: $\nu_a$=5 and $\nu_a$=6, respectively (see Table I). In these 
calculations we needed a fairly extended number of basis functions in 
the expansion (\ref{eq:expn}) for convergence. However, this is quite 
understandable because of the large energy of the initial state. 
For example, in the calculation of rotational 
transitions from that level with the following rotational indices 
$j_1=2, j_2=2$, i.e. $\nu_a=6$ and $\varepsilon^{HD}_{j_1j_2}$=536.4 
cm$^{-1}$ (Table I) all lower lying rotational levels have to be 
included in the computation.

It is known that  a quantum-mechanical
transition probability $P_{\alpha \rightarrow \beta}$ between any 
two quantum states, for example, the initial $\alpha=(j_1j_2)$ and 
final $\beta=(j'_1j'_2)$,
is inversely proportional to the energy gap $\Delta \varepsilon_{\alpha \beta}$ 
between these two states. 
In turn, the cross sections $\sigma_{\alpha \beta}(E)$ and corresponding thermal rate
coefficients $k_{\alpha \beta}(T)$
are directly proportional to the quantum probabilities. Thus
\begin{equation}
k_{\alpha \rightarrow \beta} (T) \sim {1}/{\Delta \varepsilon_{\alpha \beta}}.
\label{eq:ratio1}
\end{equation}
However, a quite unexpected result 
relating the rates in the following reactions is seen in the 
present study:
\begin{eqnarray}
&&\mbox{HD}(0)+\mbox{HD}(2) \rightarrow \mbox{HD}(0)+\mbox{HD}(1),
\label{eq:x1}\\
&&\mbox{HD}(0)+\mbox{HD}(2) \rightarrow \mbox{HD}(1)+\mbox{HD}(1). 
\label{eq:x2}
\end{eqnarray}
It is seen from Fig. \ref{fig7}, that process (\ref{eq:x1}) has much larger
 thermal rate coefficients
than process (\ref{eq:x2}). The difference between these rates is about an
order of magnitude. At the same time the energy difference between the HD molecule 
rotational states
in (\ref{eq:x1}) is $\Delta \varepsilon_{02-01}$=178.8 cm$^{-1}$,
which is larger than in (\ref{eq:x2}), for which $\Delta \varepsilon_{02-11} $=89.4 cm$^{-1}$. 
In accord with the relationship (\ref{eq:ratio1}) one could expect that
the process (\ref{eq:x1}) would have lower values of the thermal rates than the process (\ref{eq:x2}),
but it does not. This happens because in (\ref{eq:x2}) both HD molecules simultaneously change
their internal states, i.e. rotational quantum numbers. Probably, this is the reason that the
process (\ref{eq:x2}) is
much slower than (\ref{eq:x1}). This result is somewhat similar to older computational 
data on the H$_2$+H$_2$ collision \cite{gatti05,renat06}, where the authors found that
the excitation process $\mbox{H}_2(0)+\mbox{H}_2(0) \rightarrow \mbox{H}_2(4)+\mbox{H}_2(2)$
has larger cross sections at larger collision energies
than the process $\mbox{H}_2(0)+\mbox{H}_2(0) \rightarrow \mbox{H}_2(4)+\mbox{H}_2(0)$.

\section{Conclusions and future work}
Currently theoretical and experimental research in the field of the 
molecular Bose-Einstein condensates  at ultracold temperatures
is increasingly gaining momentum \cite{ni08,jin2011,shuman2010}. For example,
in the recent work \cite{shuman2010} the authors
develop a  promising approach for laser cooling of diatomic polar molecules.
The method should allow the production of large samples of molecules at ultracold 
temperatures.
Only a few of the possible practical and technological applications
where new results of this research could be used have been briefly outlined
in the Introduction.
Researchers in this new field of atomic, molecular and optical
physics have had tremendous success within last two decades. 
It is useful to have exact, high quality  full space
PESs for such polar molecule interactions. In turn the HD+HD system  could be
a prototype collision between two polar molecules with a high quality full space four-atomic PES.
In this work we performed a detailed quantum-mechanical
study of the state-resolved rotational excitation/de-excitation collisions between hydrogen molecules.
The $\mbox{HD}+\mbox{HD}\rightarrow \mbox{HD} + \mbox{HD}$ and
$\mbox{H}_2 +\mbox{H}_2 \rightarrow \mbox{H}_2 + \mbox{H}_2$ collisions
have been considered and their rotational
state-selected integral cross sections have been computed for a wide range of temperatures, i.e.
from ultracold $T \sim$ 10$^{-8}$ K to up to $T\sim $ 14000 K. We have
demonstrated that a small change in the H$_2$-H$_2$ PES to adjust for HD-HD can lead to
substantial differences in the scattering outputs, i.e. in the integral state resolved cross sections.
This calculation was carried out within a single H$_4$ PES from Ref. \cite{booth02}.

Further, in connection with the problems of coherent control of the atomic and molecular interactions
the authors of Ref.  \cite{gong2003} performed a numerical investigation of the quantum entanglement
for the case of a
nonreactive ultracold collision between two indistinguishable (H$_2$+H$_2$)
molecules. Similarly, the initial state quantum entanglement coupling
has been considered for the case of ultracold collision between identical two-atomic polar 
molecules \cite{renat2011}.
It would also be useful to mention here that
universal relations for strongly correlated fermions have been derived recently 
\cite{braaten2012,tan2008}.
Because the HD molecules are fermions with a well known
interaction potential, they could be useful, for example, in direct numerical 
verification of these 
universal relations.

The authors of Ref.  \cite{bohn2011} formulated a time-independent 
quantum-mechanical formalism to describe the dynamics of molecules with permanent 
electric dipole moments in a two-dimensional confined geometry such as in a  
one-dimensional optical lattice.  It would be useful in future investigation to 
adopt these techniques  and apply them to a fermionic 
molecule system such as HD+HD with a well-known potential \cite{booth02,robert}. It 
would be  interesting to see differences in the quantum dynamics between 
the state-resolved HD+HD rotational thermal rate coefficients of the current work 
and possible new thermal rates for the same system but when embedded in a 
one-dimensional optical lattice or a microwave trap.

\acknowledgments
This work was partially supported by Office of Sponsored Programs (OSP) of 
St. Cloud State University, and CNPq and FAPESP of Brazil.

\end{document}